\documentclass[journal]{article}
\usepackage{amsmath,color,graphicx,amssymb}
\usepackage{ifthen}
\usepackage{textcomp} 
\usepackage{gensymb}
\usepackage{placeins}
\usepackage{float}
\usepackage[bookmarks=false]{hyperref}
\usepackage{tabularx}
\usepackage{epstopdf,multirow}
\usepackage{xcolor}
\usepackage{authblk}
\begin{document}

\pagenumbering{gobble}

\providecommand{\keywords}[1]
{
  \small	
  \textbf{\textit{Keywords---}} #1
}

\date{}

\title{Improved Techniques for the Surveillance of the Near Earth Space Environment with the Murchison Widefield Array}
\author[1,2]{Brendan~Hennessy}
\author[1]{Steven~Tingay}
\author[1]{Paul~Hancock}
\author[2]{Robert~Young}
\author[1]{Steven~Tremblay}
\author[1]{Randall~B.~Wayth}
\author[1]{John~Morgan}
\author[1]{Sammy~McSweeney}
\author[1]{Brian~Crosse}
\author[1]{Melanie~Johnston-Hollitt}
\author[3]{David~L.~Kaplan}
\author[4]{Dave~Pallot}
\author[1]{Mia~Walker}
\affil[1]{International Centre for Radio Astronomy Research, Curtin University, Bentley, WA 6102, Australia}
\affil[2]{Defence Science and Technology Group, Edinburgh, SA 5111}
\affil[3]{Department of Physics, University of Wisconsin--Milwaukee, Milwaukee, WI 53201, USA}
\affil[4]{International Centre for Radio Astronomy Research, University of Western Australia, Crawley, WA 6009, Australia}

\maketitle

\begin{abstract}	
In this paper we demonstrate improved techniques to extend coherent processing intervals for passive radar processing, with the Murchison Widefield Array. Specifically, we apply a two stage linear range and Doppler migration compensation by utilising Keystone Formatting and a recent dechirping method. These methods are used to further demonstrate the potential for the surveillance of space with the Murchison Widefield Array using passive radar, by detecting objects orders of magnitude smaller than previous work. This paper also demonstrates how the linear Doppler migration methods can be extended to higher order compensation to further increase potential processing intervals.
\end{abstract}


\section{Introduction}
\label{sec:into}
The Murchison Widefield Array (MWA) is a radio telescope located in Western Australia~\cite{2013PASA...30....7T}. It is the low frequency precursor to the upcoming Square Kilometre Array\cite{2015IAUGA..2252814B}. Operating in the frequency range 70 $-$ 300 MHz, the main scientific goals of the MWA are to detect radio emission from neutral hydrogen during the so-called Epoch of Reionisation (EoR), to study our Sun and heliosphere, the Earth's ionosphere, and radio transient phenomena, as well as map the Galactic and extragalactic radio sky~\cite{2013PASA...30...31B}. The MWA is sensitive to, as a source of radio interference, terrestrial transmissions, such as FM radio and digital TV, reflected by objects in low Earth orbit~\cite{2013AJ....146..103T}, but as far as the Moon in the case of a global ensemble of transmitters~\cite{2013AJ....145...23M}. Recently, it has been shown that passive radar techniques using the MWA can detect and range these objects in low Earth orbit (LEO), allowing orbits to be generated~\cite{7944483}.

With the ever increasing number of human-made objects in Earth orbit, the reduction in barriers and costs of putting an object in orbit, and the rapid uptake of small-satellite technology, the surveillance of space is an increasingly important area of interest. This is highlighted by the increasing fears over the Kessler Syndrome, a scenario in which the density of debris in LEO is high enough that a collision causes a chain reaction of subsequent collisions, potentially rendering near-Earth space inaccessible~\cite{7024495}. 

Typically, the radars employed for the surveillance of space are very narrowly focused tracking radars, and are only able to surveil a small solid angle at any one time~\cite{NAP13456}. The MWA has many beneficial characteristics for passive radar: the wide-area field of view (100s to 1,000s of square degrees); its location at the Murchison Radio-astronomy Observatory (MRO), a designated radio quiet zone (subject to very low levels of interference); and the MWA's high sensitivity across a wide frequency range coinciding with numerous terrestrial transmitters.

Recent publications detailing space debris detection with radar, both passive and active, highlight the need to incorporate the object's trajectory into the processing in order to enable longer coherent processing intervals in order to detect smaller objects~\cite{5494621}\cite{7002727}\cite{7490614}.

In this paper we build upon previous work to further develop processing strategies using the MWA as an element in a passive radar system, particularly, in extending coherent processing intervals. In Section \ref{Sec:prior} prior work is detailed, including recent work in upgrading the MWA. Section \ref{Sec:proc} covers the processing strategies generally used in extending processing intervals in passive radar, as well as the specific techniques used in this paper. Section \ref{Sec:results} covers a 2016 observational campaign and details some results demonstrating improvements in the detection of space debris. Last, Section \ref{Sec:future} details future directions in this area and includes plans for future observational campaigns.

\section{Prior work}
\label{Sec:prior}
The MWA originally consisted of 128 `tiles', with each tile made up of 16 dual-polarised wide-band dipoles in a 4x4 configuration
. The MWA covers a frequency range of 70 MHz to 300 MHz, with an instantaneous bandwidth of 30.72 MHz.

The MWA has previously been used to demonstrate non-coherent\footnote{That is, without reference to the transmitted signal.} detections of the Moon as well as the International Space Station (ISS) using reflected FM radio~\cite{2013AJ....145...23M}\cite{2013AJ....146..103T}. This work went further to present simulated results predicting that the MWA is capable of detecting much smaller debris-sized objects\footnote{Debris radius of $> 0.5 m$ to $∼1,000 km$ altitude~\cite{2013AJ....146..103T}.}.

Following the non-coherent detection of the ISS, a data collection was undertaken in 2015 in order to demonstrate the use of the MWA with passive radar, focusing on the ISS. The MWA, despite being in the MRO, was able to directly receive the reference signal from Geraldton, the FM signal diffracting the three hundred kilometres. By comparing the directly transmitted signal, observed at the horizon, and the reflected surveillance signal, passive radar techniques were used to detect aircraft, ionised meteor trails and the ISS~\cite{7944483}.

This work proceeded to show that including the bistatic-range and Doppler measurements of the ISS greatly improves the ability to generate an orbit, in this case from a single pass. This is especially notable as this was achieved with a single 10 kW radio station, at an elevation of over 60\degree  ~at the point of closest approach, well outside of the primary beam of the transmitter.

\begin{figure}[ht!] 
\begin{center}
\includegraphics[width=\columnwidth]{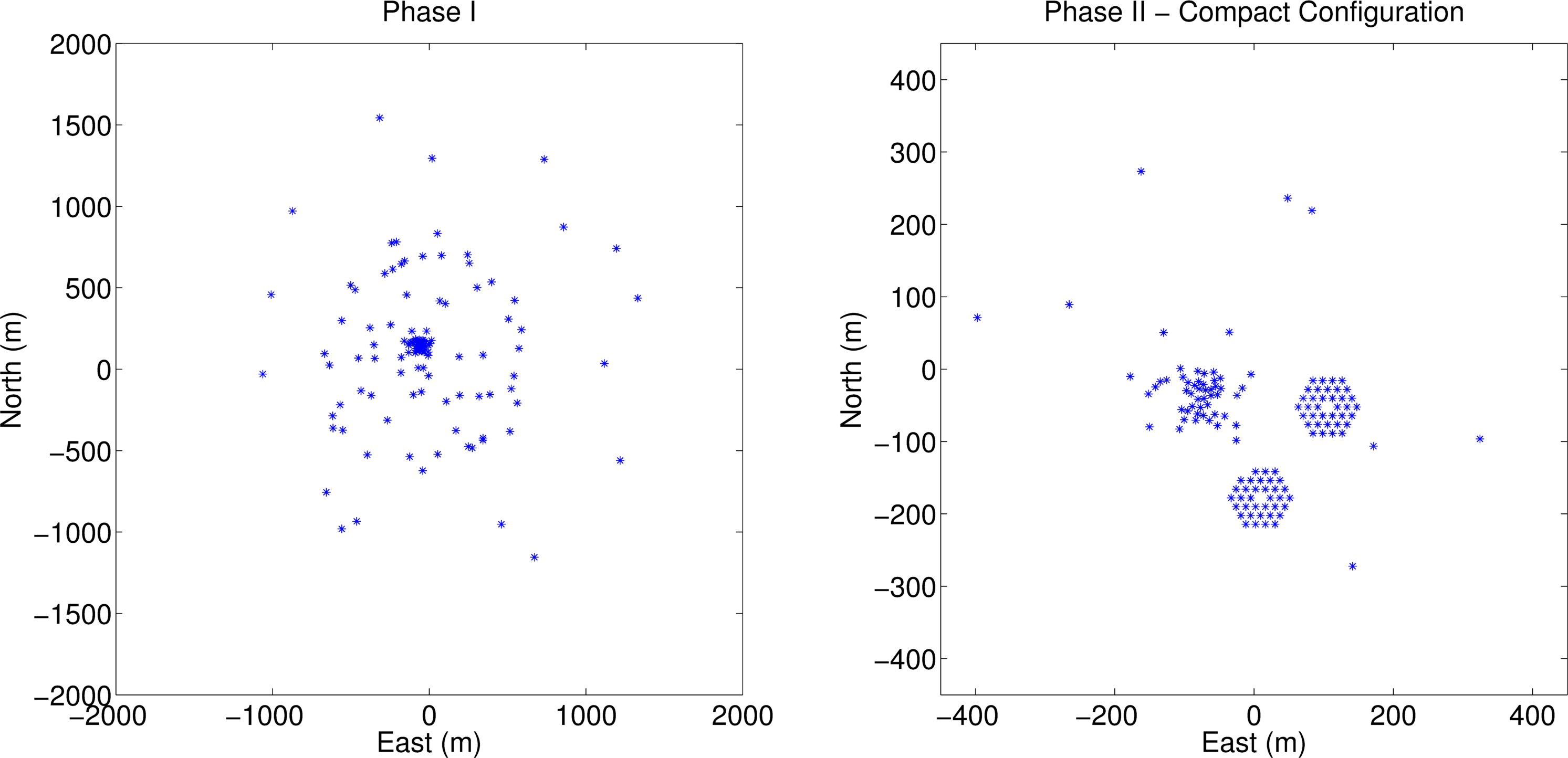}
\end{center}
\vspace{-2ex}
\caption{The left section shows the Phase I array. The right section shows the Phase II compact configuration array.}
\label{fig:phase1vphase2}
\end{figure}

The MWA has recently undergone an upgrade, from Phase I to Phase II, doubling the number of tiles, allowing the array to be reconfigured between `extended' and `compact' configurations~\cite{pase22018article}. Figure \ref{fig:phase1vphase2} shows both the Phase I array layout as well as the Phase II compact configuration array layout, including the two compact hexagons.

Observation data are collected through the Voltage Capture System for recording high time and frequency resolution data~\cite{2015PASA...32....5T}. These voltages are collected after two polyphase filter bank (PFB) stages which critically sample the data into coarse 1.28 MHz channels and then 10 kHz fine channels. These data are phase-adjusted to account for the cable delays and then a further calibration solution is applied to remove instrumental and atmospheric effects.
The calibration solutions are produced by recursively accounting for residuals after removing the visibilities from known strong compact sources~\cite{2013PASA...30....7T}. 

The second PFB stage is then inverted to combine the 10kHz sub-channels into timeseries data, reconstituting the FM band for beamforming and delay-Doppler processing.

\section{Processing Strategies}\label{Sec:proc}
Space debris radar research consistently highlights the need for much longer processing intervals in order to improve system sensitivity; this mirrors similar considerations in passive radar. With the increase in computing power and available memory, extending processing intervals is far more achievable. This raises challenges for high-speed and manoeuvring targets as increasing the coherent processing interval (CPI) will lead to range and Doppler migration. That is, the target may be moving sufficiently fast to `smear' returns across multiple delay and Doppler bins during a single CPI, thereby reducing the target's power in each delay and Doppler bin. Traditionally, this has meant constraining CPIs to small values. Doppler migration is further exacerbated by increasing CPIs, as the Doppler resolution is proportional to the CPI length.

Mirroring the need to incorporate debris trajectory to improve performance, as in Section \ref{sec:into} above, incorporating target trajectory to avoid range and Doppler migration is a consistent theme in general passive radar research~\cite{6869185}\cite{4148608}.

The classic solution for handling range migration is the Keystone Transform. The Keystone Transform is a frequency dependant slow time resampling, to decouple range and Doppler, removing all linear range migration~\cite{745691}. 
In order

to extend CPIs for detecting high-speed and manoeuvring targets, ambiguity surface processing has been extended to incorporate acceleration~\cite{4653940}\cite{7497376}.
As shown in \eqref{eq:ddaccel}, given a surveillance signal $s(t)$ and reference signal $r(t)$, the narrowband ambiguity function is a matched filter over CPI $T$ to delay $\tau$ and Doppler $v$, but also extended to include the Doppler rate $w$. The rate of Doppler change is analogous to target acceleration.
\vspace{-1ex}
\begin{equation}
\label{eq:ddaccel}
    \chi(\tau,v, w) = \int_T s(t)r^*(t-\tau)e^{-j2\pi(vt + \frac{1}{2}wt^2)}\,dt
\end{equation}

This processing only removes Doppler migration due to acceleration.  Range migration caused by accelerating targets will not be compensated as the delay term $\tau$ is not adjusted. This method also requires the delay Doppler map to be recomputed for each acceleration hypothesis. Because evaluating the complete ambiguity function is computationally very expensive, approximations to these methods are used~\cite{1459145}~\cite{Palmer:2011:OIO:2016157.2016196}.

These involve producing a range-compressed pulse stack and resolving Doppler with a Fourier transform. As well as being far more efficient to produce, they also enable more efficient methods of detecting accelerating targets.

In recent years the most common method for detecting accelerating targets is the Fractional Fourier Transform~\cite{TRANROCCOSANDUN2}. The Fractional Fourier transform is a generalisation of the classic Fourier transform to an arbitrary order, or `angle', in the time and frequency plane, with the Fourier transform representing a $\frac{\pi}{2}$ order transform. For a target undergoing linear acceleration, the Doppler returns will be smeared when extracted with the Fourier transform. However, with a suitably chosen angle, the returns will be localised to a single bin with the Fractional Fourier Transform.

A novel, and simple, improvement has come from noting that a target undergoing linear acceleration will produce a linear frequency modulation response in slow-time.

The Fractional Fourier transform decouples the frequency-time dependence in the signal to produce a tone. This leads to a great simplification.  Rather than using the Fractional Fourier transform, a non-linear phase correction can be applied to dechirp the motion-induced chirp, and then the Fourier transform can be used to resolve Doppler as before~\cite{TRANROCCOSANDUN1}.

The non-linear phase correction, across slow-time, is given by:
\begin{equation}
\label{eq:dechirp}
e^{-2 j \pi c_{r} t^2},
\end{equation}
where the dechirp rate $c_{r}$ is given by $\frac{a}{\lambda}$, $a$ is the acceleration hypothesis and $\lambda$ the wavelength.

This is a much more efficient method for detecting accelerating targets. Not only can it be incorporated into approximations to the complete ambiguity surface, but significantly, the delay-Doppler map does not need to be recomputed for acceleration hypotheses.  The compressed pulse stack can be computed once, and then only the Doppler-resolving Fourier transform needs to be repeated for each acceleration value.

\begin{figure}[ht!]
\begin{center}
\includegraphics[width=\columnwidth]{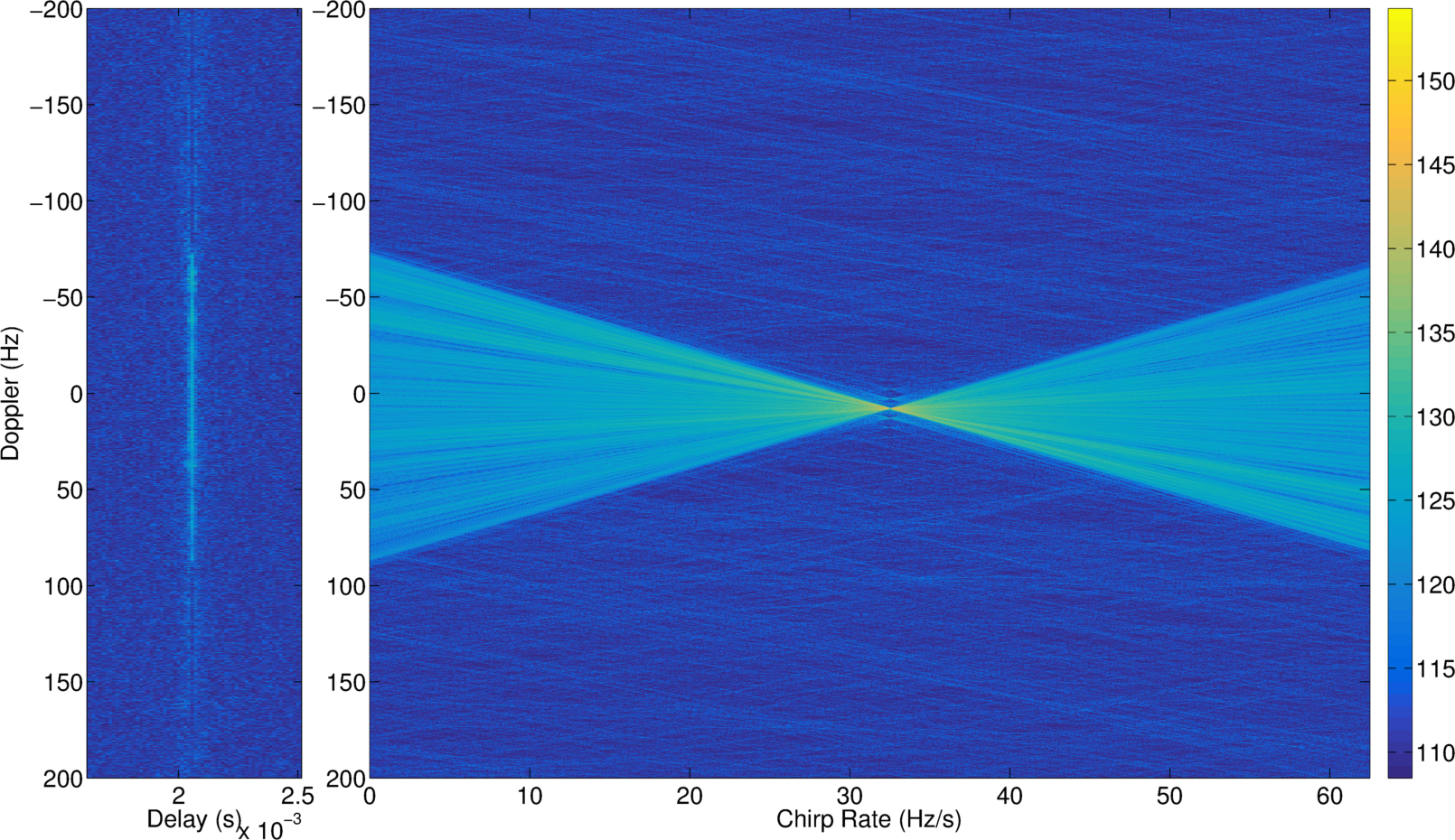}
\end{center}
\vspace{-2ex}
\caption{Ambiguity surface signal power in dB of the ISS, processed with a five second CPI, along with the ISS returns processed with different dechirp hypotheses.}
\label{fig:full-dechirp}
\end{figure}

An example of this is shown in Figure \ref{fig:full-dechirp}, showing the results of a five second CPI for the ISS moving through zero Doppler. The left panel shows a region of the delay Doppler map with the ISS manifesting as a vertical line, smeared across hundreds of Doppler bins, the result of Doppler migration. The right shows a single range bin reprocessed for a range of dechirp hypotheses. The target SNR is increased when the target is localised to a single Doppler frequency, with the appropriate dechirp rate.

A common approach is to handle range and Doppler migration separately, rather than attempting to compensate for both in one transform~\cite{TRANROCCOSANDUN3}\cite{7914094}. Traditionally, this is achieved by processing to remove linear range migration and then processing to remove Doppler migration due to linear acceleration. The efficacy of extending processing intervals is limited in this case, as a target undergoing acceleration will result in non-linear range migration. 

Orbital kinematics, however, are ideal for this type of extended integration processing, as orbital object motion is very stable and reliable. In the space situational awareness (SSA) case, the acceleration that is detected is primarily due to apparent radial acceleration caused by the changing bistatic geometry.

For an object in orbit, the major contribution to acceleration is Earth's gravity, combined with other much smaller forces such as atmospheric drag and space weather effects. The bistatic acceleration, rather Doppler rate, detected in bistatic radar processing will be dominated by the relative geometry, as the Doppler rapidly changes as the object passes overhead. This is incredibly beneficial to extended processing, as the range and Doppler migration effects are quite separate. The rate of Doppler change will be highest when the Doppler is zero, at the object's closest point, transitioning from positive to negative Doppler. Therefore Doppler migration is greatest when the range migration is zero. Conversely the range migration will be at its maximum at larger ranges, at which the Doppler magnitude is at a maximum, and Doppler rate  is approaching zero.

This only applies to objects in relatively stable orbits; if an object was falling directly toward the radar then the true acceleration, due to gravity, would dominate the Doppler rate, and these methods would not be suitable. Two stage linear range and Doppler-rate methods will tend to defocus returns of non-orbital objects, as seen in returns from meteors and aircraft.

\section{Results}
\label{Sec:results}
Previously reported detections of the ISS with passive radar using the MWA were achieved by measuring range, Doppler, azimuth, and elevation and then inferring azimuth, and elevation rates~\cite{7944483}. The results included in this paper also measure Doppler rates using the dechirping method, mentioned earlier. However, azimuth and elevation rates are also directly measured by constructing the surveillance signal based on ephemerides. The surveillance signal is constructed by adjusting the beamforming weights during the CPI so that the target is tracked throughout. This sub-CPI beamforming adjustment, whilst suitable for demonstrating improved detection performance, is not tractable for wide-field, blind searches as it is essentially adding extra dimensions to the ambiguity function.

\begin{figure}[ht!]
\begin{center}
\includegraphics[width=\columnwidth]{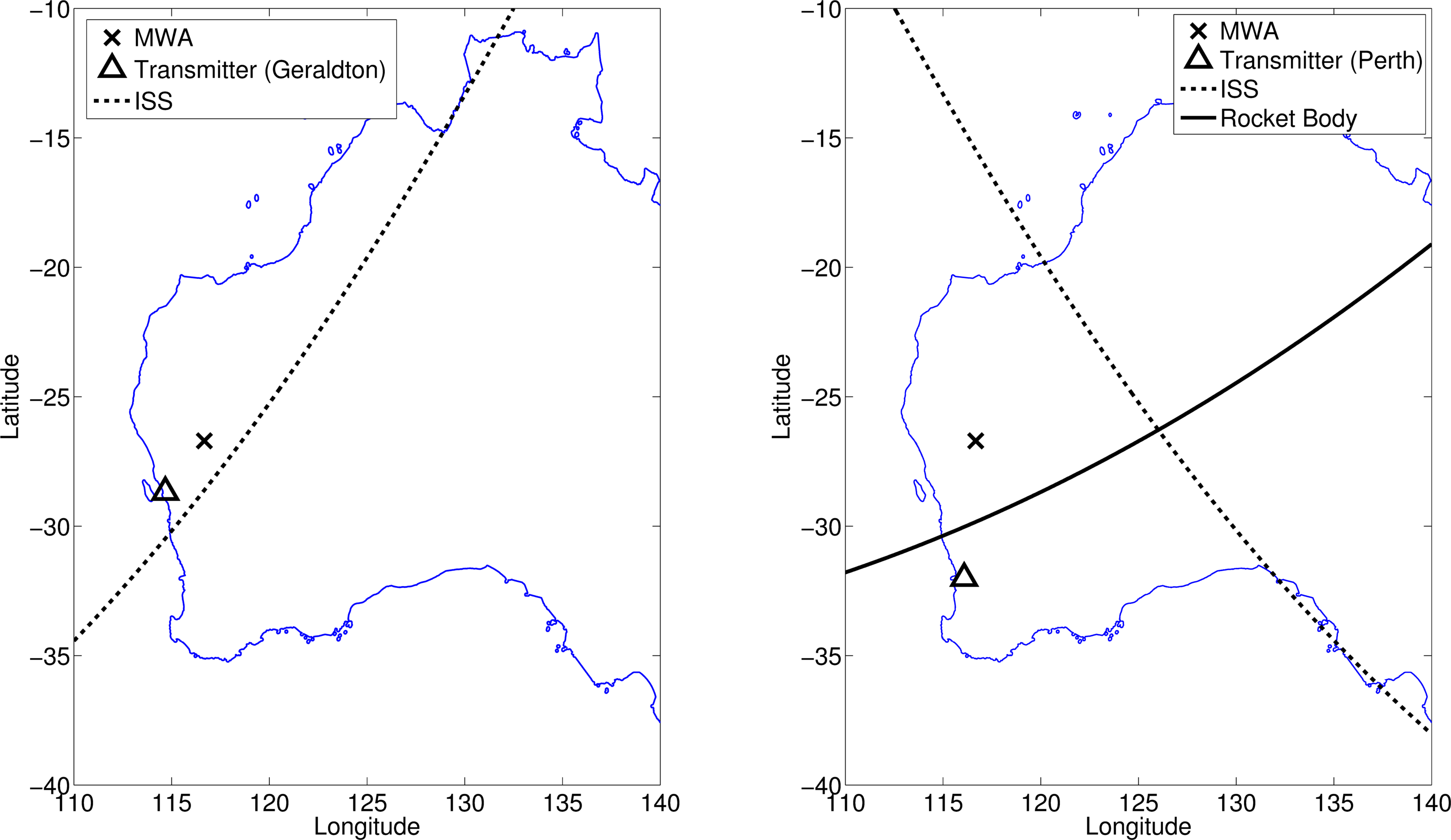}
\end{center}
\vspace{-2ex}
\caption{The 2015 (left) and 2016 (right) data collection configurations, overlaid on a map of Western and Central Australia.}
\label{fig:MAPofTLEs}
\end{figure}

\subsection{2015 Data Reprocessing}
The first passive radar detections of an object in orbit, were generated from a dedicated MWA observation collected in 2015~\cite{7944483}. In this 2015 dataset the ISS passes almost directly above the MWA\footnote{Rather, passing at a maximum elevation of 70\degree\space from the MWA.}, providing ideal conditions for detection. However, a significant issue with this earlier work was the ISS’ SNR, or detectability, decreasing with longer processing intervals. 

This was due to the ISS' return smearing through the ambiguity surface and search parameters. As well as the Doppler and range smearing, additional migration occurred in beamforming direction and direction rates. This is due to the high angular resolution achievable with the MWA, particularly with the long baselines in the initial configuration. With an angular resolution less than one tenth of a degree, our standard processing could not coherently follow the ISS as it subtends almost three degrees per second at the point of its closest approach.

Another issue was that the reference signal was formed directly from the MWA observation data. This had the result of limiting potential baseline lengths; with Geraldton being 300 km away the elevation of the ISS from the transmitter was large enough that the ISS was not in the primary transmitter beam. The reference signal is likely to have suffered from multipath effects, diffracting over such a distance, which may raise the clutter floor or cause destructive interference. The map of the collection geometry with the transmitter in Geraldton, the MWA and the ISS path is shown in Figure \ref{fig:MAPofTLEs}. The MWA was in its Phase I configuration, as shown in Figure \ref{fig:phase1vphase2}.

\begin{figure}[ht!]
\begin{center}
\includegraphics[width=\columnwidth]{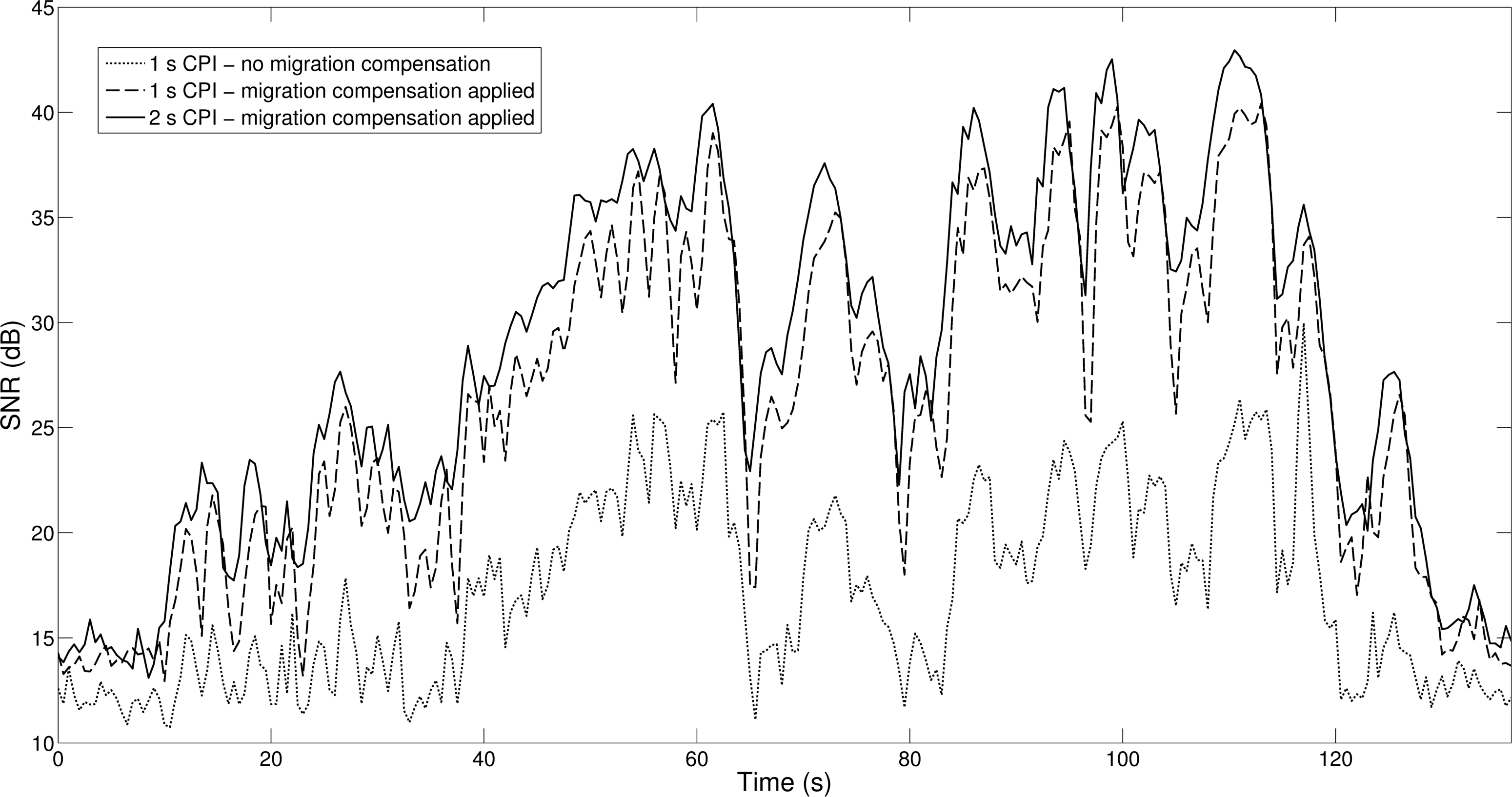}
\end{center}
\vspace{-2ex}
\caption{The SNR (dB) of the ISS with no migration compensation applied (corresponding to initial results~\cite{7944483}) through to full range and Doppler migration compensation.}
\label{fig:migration_neat}
\end{figure}

Figure \ref{fig:migration_neat} shows the reprocessing of the ISS' pass from the 2015 data using the Keystone Transform to mitigate range migration as well as dechirping to mitigate Doppler migration; for comparison it also shows the SNR 
from the original publication, without any migration compensation.

The migration compensation methods result in a dramatic increase in the SNR due to the improved processing gain. In this instance the ISS is now initially detected at a slant range of over 1,000 km from the MWA. More importantly, however, the SNR of the ISS increases with the CPI, with Figure \ref{fig:migration_neat} showing an expected \texttildelow 3dB increase changing from a one second CPI to a two second CPI.

\begin{figure}[ht!]
\begin{center}
\includegraphics[width=\columnwidth]{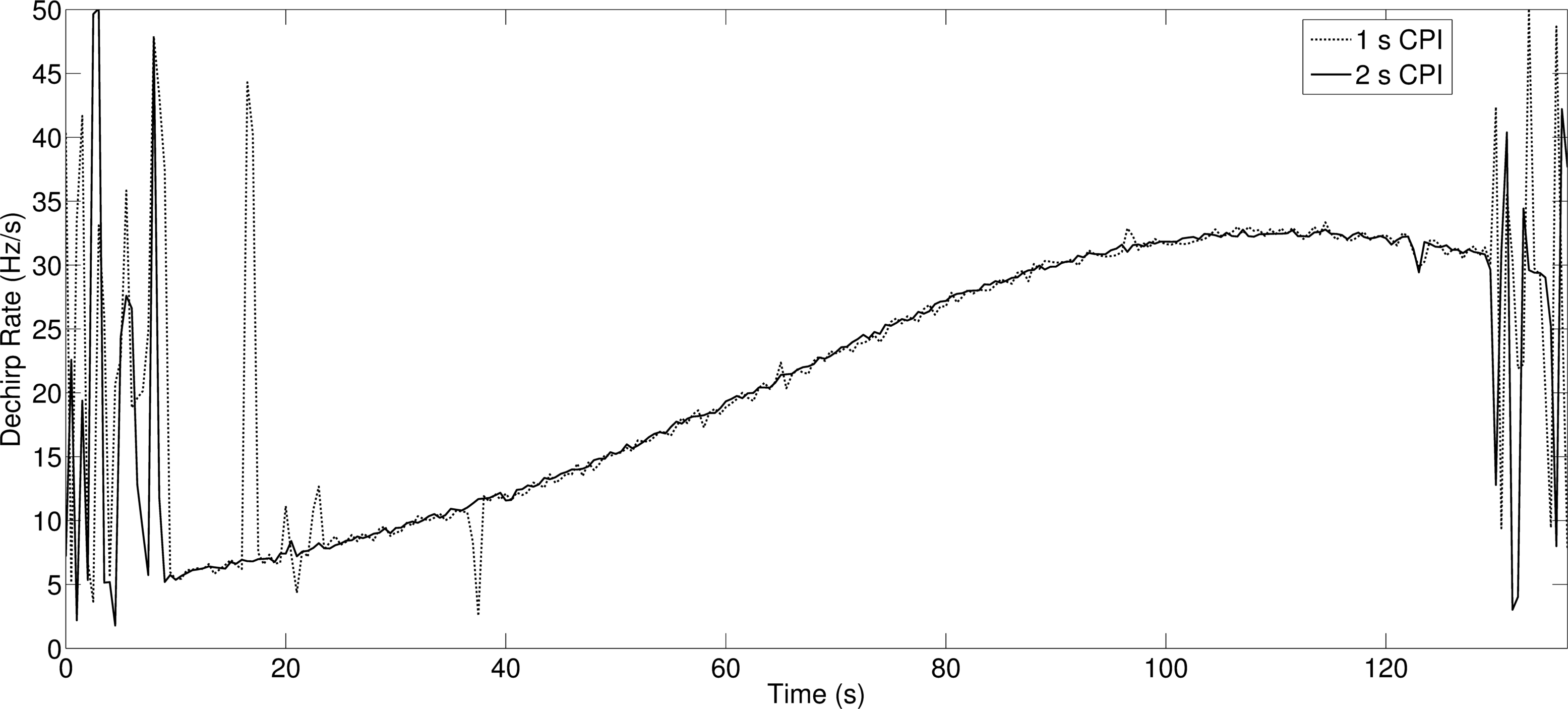}
\end{center}
\vspace{-2ex}
\caption{Detected dechirp rate corresponding to the SNR in Figure \ref{fig:migration_neat}.}
\label{fig:doppler_neat}
\end{figure}

Figure \ref{fig:doppler_neat} shows the associated chirp rates for the SNR returns in Figure \ref{fig:migration_neat}. It shows a very clear trackable curve, and again, results improve as the CPI increases.

\subsection{2016 Data Collection}
As part of a broader demonstration campaign, FM band collections were undertaken by the MWA in late 2016,  targeting the ISS again, and also some lower Radar Cross-Section (RCS)  objects~\cite{2017amos.confE..63M}. Specifically, the MWA focused on a rocket body, a large piece of debris in LEO. For this collection, a reference signal was recorded separately in Perth, 600 km away from the MWA. The reference collection recorded the entire FM band; the main focus was three 100 kW omnidirectional, mixed polarisation stations from the Bickley transmitters. There was no direct-path FM signal present in the MWA collections, due to the specific ducting/propagation conditions for the particular analogue beamforming configuration.
The particular collection geometry, shown in Figure \ref{fig:MAPofTLEs}, was far from ideal, being so far from the MWA. The MWA was in the Phase II compact configuration, as shown in Figure \ref{fig:phase1vphase2}.

\begin{figure}[ht!]
\begin{center}
\includegraphics[width=\columnwidth]{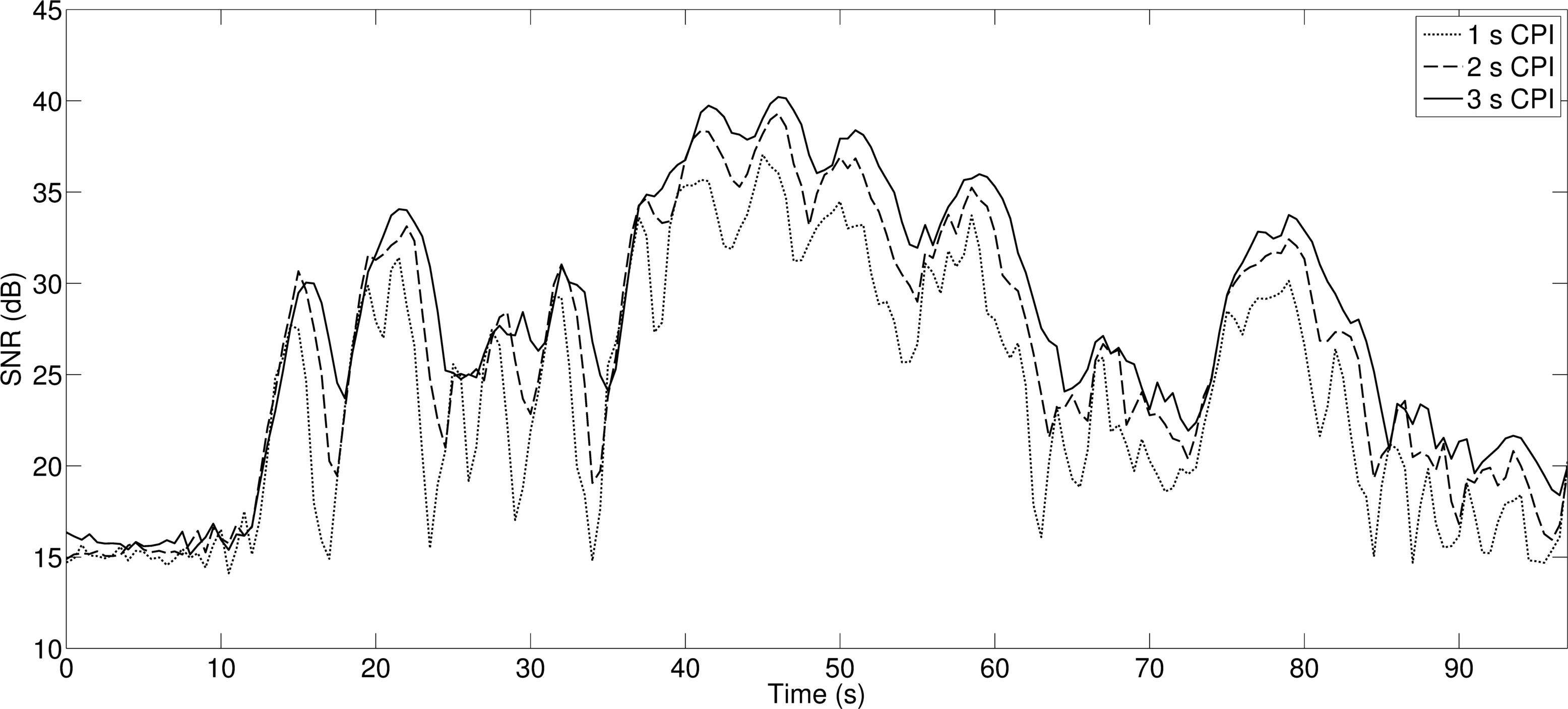}
\end{center}
\vspace{-2ex}
\caption{The SNR (dB) of the ISS from the 2016 data collection with CPI lengths of one, two and three seconds. The SNR improves with CPI.}
\label{fig:ISS_SNR_CAP2}
\end{figure}

Figure \ref{fig:ISS_SNR_CAP2} shows the SNR of the ISS for three different CPI lengths, with the migration compensation techniques applied. It shows the ISS being detected with significant SNR despite being so far from Perth and the MWA. The closest approach was at a bistatic range of 1,436 km, with a total reflected path distance never less than 2,000 km. The ISS was almost certainly in the main beam of the transmitter, being less than 15\degree\space elevation from Perth for the duration of the observation.

An interesting aspect of these passes was that the reference signal was inadvertently recorded without any clock synchronisation. In order for the reference signal to be used for coherent processing, it needed to be synchronised, both in time and frequency. Bright meteor returns were used for coarse time-synchronisation, and then the returns of the ISS itself were used for fine time and frequency alignment. That is, reflections from meteor trails are sufficiently large that they are easily detected incoherently, such that a time-aligned MWA surveillance signal, known to contain meteor returns, can be formed. This meteor reflection is correlated across the reference signal until the coherent meteor return is detected, providing a coarse time alignment (as the exact location of the meteor return is not known). This coarse time alignment then allows a surveillance beam to be formed towards the ISS, so that coherent ISS detections, in delay and Doppler, can then be compared to expected returns from ephemerides to provide fine time and frequency offsets, which are corrected. This time and frequency alignment was sufficient to be able to detect the other target from this collection, a rocket body.

\subsection{Rocket Body}
The rocket body was a stage from the Atlas-Centaur launch system, launched in August 1972. The object is 9 m long with a diameter of 3.05 m. Based on a perfectly conducting cylinder of the same dimensions, the RCS at 100 MHz is estimated to be between $10 m^2$ and $100 m^2$, an order of magnitude or more smaller than the ISS~\cite{6651980}.

\begin{figure}[ht!]
\begin{center}
\includegraphics[width=\columnwidth]{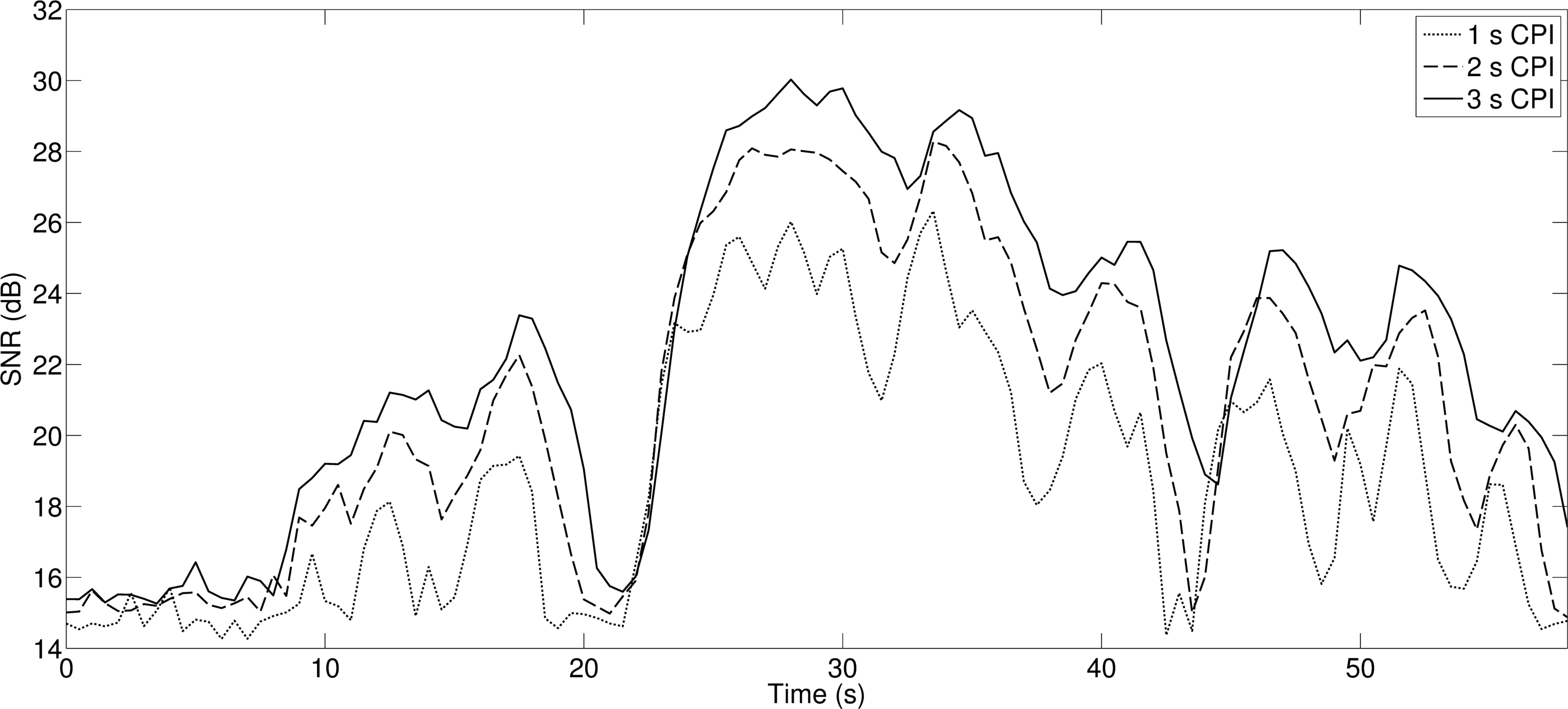}
\end{center}
\vspace{-2ex}
\caption{The SNR of the Atlas-Centaur rocket body from the 2016 data collection with CPI lengths of one, two and three seconds. The SNR improves with CPI.}
\label{fig:ATLAS_SNR_CAP2}
\end{figure}

As shown in Figure \ref{fig:MAPofTLEs}, this pass is sub-optimal for passive radar processing, as the object passes much closer to the transmitter than the receiver. At the point of closest approach, at a bistatic range of 894 km, the rocket body was at an elevation of 73\degree\space from the transmitter; well outside the transmitter's main beam. Figure \ref{fig:ATLAS_SNR_CAP2} shows the SNR of the rocket body for this pass when processed using the migration compensation techniques. Unlike the ISS, the rocket body is not detectable without accounting for the range and Doppler smearing. These migration techniques allowed coherent improvements in detectability with CPIs up to 10 s, at some points in its trajectory.

These results, as well as the sub-optimal geometry, both suggest that much smaller targets will be readily detectable. The best SSA results will be achieved when surveilling the wide area directly above the MWA, when the objects are illuminated by the main beam of FM radio transmitters.

\subsection{Higher Order Hypotheses}
The modified narrowband ambiguity function \eqref{eq:ddaccel} has previously been extended to incorporate higher order Doppler migration compensation terms~\cite{4653940}. This still requires recomputation of the ambiguity surface, so is very computationally expensive. Similarly, the dechirp processing \eqref{eq:dechirp}, can be extended to include non-linear and higher order frequency modulation terms. With this extension the non-linear phase adjustment can be applied to dechirp and dejerk the target's Doppler migration.

\begin{equation}
e^{-2 j \pi (c_{r} t^2 + c_{j} t^3)}
\end{equation}

The chirp rate, now varying over time, is given by $c_{r} + c_{j}t$, to include $c_{j}$, the rate of change of the chirp rate. This rate is here referred to as the `jerk', as it is analogous to the rate of change of bistatic acceleration.

\begin{figure}[ht!]
\begin{center}
\includegraphics[width=\columnwidth]{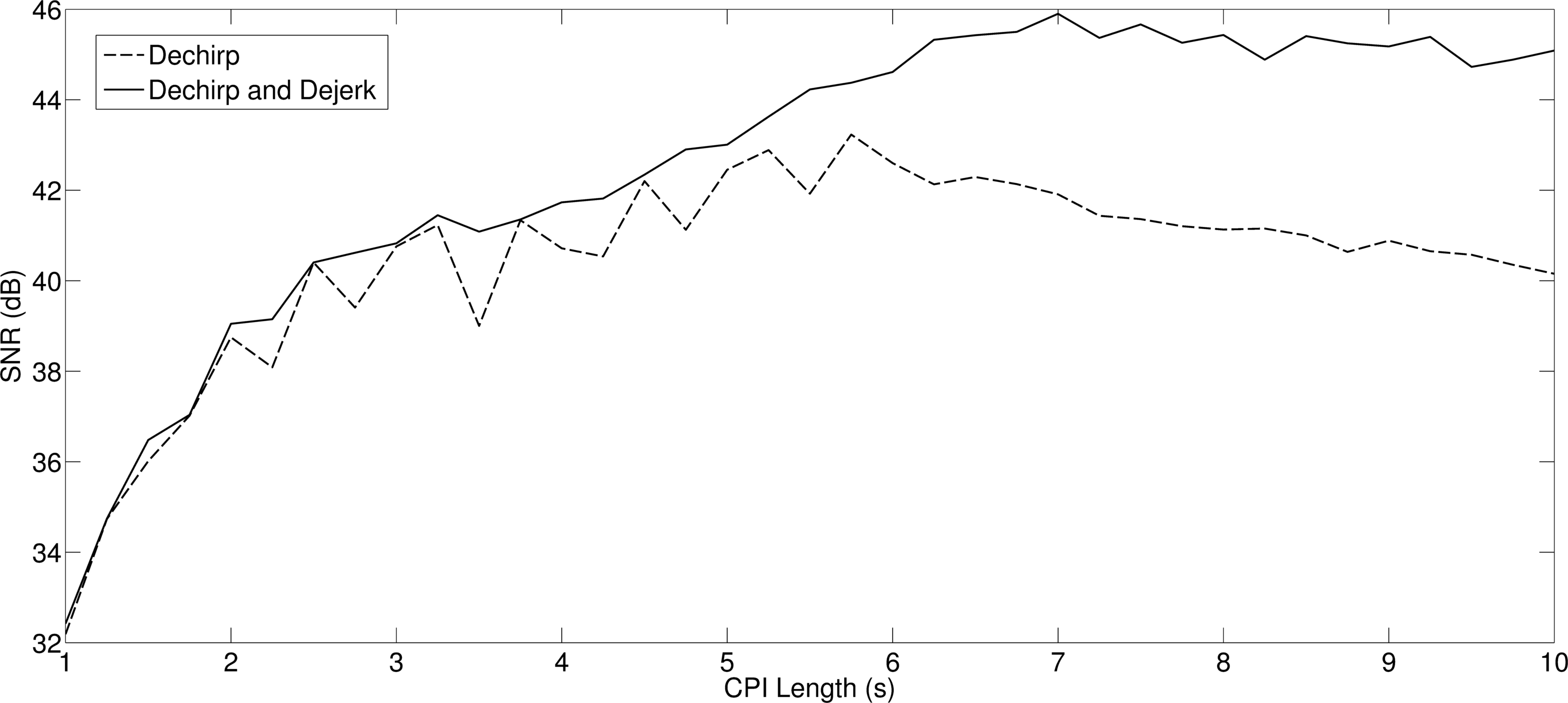}
\end{center}
\vspace{-2ex}
\caption{A single snapshot in time of the SNR of the rocket body and how it varies with CPI, comparing the dechirp technique with the higher order dechirp and dejerk.}
\label{fig:SNR_jerks}
\end{figure}

Figure \ref{fig:SNR_jerks} shows a single snapshot in time of the SNR of the rocket body
and how it varies with CPI. It shows that by incorporating the higher order motion compensation term, at least in Doppler, a modest improvement in maximum CPI length is achieved. However, like previous methods mentioned, higher order compensation means further extending the search space processing, as required for the detection of unknown targets.

\section{Future Work}
\label{Sec:future}
In order to conduct surveillance processing with the MWA, more tractable processing strategies will need to be developed. The methods in Sections \ref{Sec:proc} and \ref{Sec:results} demonstrate improvements in the performance of passive radar using the MWA.  However, applied naively they would result in a 10-dimensional ambiguity surface to search over. An improvement would be to work backwards from the orbital parameters of interest, as an object's orbit will largely constrain and define most of the other search parameters~\cite{8448187}. Incorporating orbital kinematics, even including highly eccentric orbits, will greatly reduce the subspace of possible values for other parameters, including range and Doppler migration factors, as well as pointing directions and their associated rates.

Further improvements to these methods would benefit from additional data collections targeting even smaller RCS objects, including space debris, passing above the MWA.  These data collections are planned for the near future and will be reported in future publications.

\section{Conclusion}
\label{Sec:conc}
This paper covers improved techniques for extending coherent processing intervals with passive radar for space surveillance using the MWA. Specifically, applying a two stage linear range and Doppler migration compensation by utilising Keystone Formatting and a recent dechirping method.

These methods have limitations for accelerating targets, but work well for objects in orbit by handling migration due to apparent radial acceleration due to the bistatic geometry. 

These methods are then used to further demonstrate the potential for the surveillance of space with the Murchison Widefield Array using passive radar, by detecting objects at least an order of magnitude smaller than previous work. Notably, the detection conditions were difficult, including the sub-optimal bistatic geometry and the separately recorded and unsynchronised reference signal.

This paper also demonstrates how linear Doppler migration methods can be extended with higher order compensation to further increase potential processing intervals.

Finally, this paper outlines approaches that may improve these techniques, by directly incorporating orbital parameters into the ambiguity surface formation. Planned future collections will further improve space debris detection and tracking with the MWA. 

\section{Acknowledgement}
The authors would like to thank Dr. Bevan Bates from the University of Adelaide and Dr. Kin Shing Yau from the Defence Science and Technology Group for their assistance towards this paper.

This scientific work makes use of the Murchison Radioastronomy Observatory, operated by CSIRO. We acknowledge the Wajarri Yamatji people as the traditional owners of the Observatory site. Support for the operation of the MWA is provided by the Australian Government, under a contract to Curtin University administered by Astronomy Australia Limited. We acknowledge the Pawsey Supercomputing Centre which is supported by the Western Australian and Australian Governments.

This research is supported by the Defence Science and Technology Group. 
\bibliographystyle{IEEEtran}
\bibliography{./paperonebib}

\end{document}